\documentclass[11pt]{article}
 
\usepackage[utf8]{inputenc}
\usepackage[english]{babel}
\usepackage{graphicx}
\usepackage{subcaption}
\usepackage{epsfig}
\usepackage{amssymb}
\usepackage{bm}
\usepackage{amsmath}
\usepackage{color}
\usepackage{hyperref}
\usepackage{times}
\usepackage{float}
\usepackage{times}

\def\be{\begin{equation}}
\def\te{\end{equation}}
\def\bea{\begin{eqnarray}}
\def\nn{\nonumber\\}
\def\tea{\end{eqnarray}}

\begin{document}
\title{Limits on quantum measurement engines}
\author{Guillermo Perna and Esteban Calzetta\\
Universidad de Buenos Aires, Facultad de Ciencias Exactas y Naturales,\\
 Departamento de Física. Buenos Aires, Argentina, and\\
CONICET - Universidad de Buenos Aires,\\
 Instituto de Física de Buenos Aires (IFIBA). Buenos Aires, Argentina\\
gperna@df.uba.ar ; calzetta@df.uba.ar}
\maketitle

\begin{abstract}
A quantum measurement involves energy exchanges between the system to be measured and the measuring apparatus. Some of them involve energy losses, for example because energy is dissipated into the environment or is spent in recording the measurement outcome. Moreover, these processes take time. For this reason, these exchanges must be taken into account in the analysis of a quantum measurement engine, and set limits to its efficiency and power. We propose a quantum engine based on a spin 1/2 particle in a magnetic field and study its limitations due to the quantum nature of the evolution. The coupling with the electromagnetic vacuum is taken into account and plays the role of a measurement apparatus. We fully study its dynamics, work, power and efficiency. 
\end{abstract} 

\section{Introduction}

The advancement of quantum measurement engines \cite{Talkner17,Talkner18,Elouard17,Elouard18,Laflamme2020,Auffeves21,Auffeves21b,Auffeves21c,Chand2017,Chand2018,Chand2021,Biswas23,Santos2023,Serra23,Goold23,Cangemi2023,Anka21} not only offers significant technological potential but also carries profound implications for fundamental physics  \cite{Auffeves19}. Therefore, it is crucial to develop precise methods for assessing their performance in terms of both efficiency and power.

In a quantum measurement engine, as opposed to a thermal one \cite{Lutz12,Lutz14,Christian,Nico22,Milton22,Milton22b,Deffner22,Lili23}, energy is introduced into the system not in the form of heat but through an interaction with an external apparatus, typically interpreted as a measurement on the system. If this interaction indeed constitutes a measurement, the resulting information could be used to influence the subsequent operation of the engine, similar to a Maxwell's demon scenario \cite{LeffRex2003,Anders17a,Anders17b,Kim18,Seah19,Jordan23}. In a simpler context, the key is that a measurement is conducted, the specific outcome being inconsequential \cite{Das18,Kosloff19}. This type of quantum measurement engine is the focus of our discussion.

While a comprehensive theory of quantum measurement remains elusive \cite{Wheeler_Zurek,Braginsky1992,Mittelstaedt98,Wiseman09,Devoret2010,Busch16,Jacobs14,Castro19,Hance22}, there is a general consensus that a measurement process comprises several stages, which can be summarised as follows:

1) The system to be measured comes into contact with the measuring device, which is usually a macroscopic body significantly larger than the system \cite{Loveridge2011,GS90}, and they evolve unitarily into an entangled state.

2) This initially entangled, yet pure state, transforms into a mixed state where distinct measurement outcomes, stored within the apparatus, can be assigned definite probabilities. If the Schrödinger equation remains unaltered, this decoherence process is attributed to the interaction of the system-apparatus complex with an external environment \cite{Walls85,JP99,JP16,Weinberg16,Lone17,Foti19,Bhaumik22,Yahalom23}.

3) The mixed state evolves into a single pure state, in which the measurement recorded by the apparatus displays a definite result \cite{ZV88,ABN06,ABN13,Nieuwenhuizen23,Horodecki17}. Since we are concerned with engines that function regardless of the measurement outcome,  the completion of this stage is not essential for our analysis.

4) If the engine is designed to operate cyclically, the recorded information must be erased before the next cycle begins.

It is evident from this overview that a quantum measurement involves repeated energy exchanges among the system, apparatus, and the broader external world, some of which are in the form of heat. For instance, in stage (4), the erasure of the apparatus's memory would entail the release of heat, as dictated by Landauer's Principle \cite{LeffRex2003}. In stage (3), the transition from a mixed to a pure state implies a definite decrease in entropy, which should also involve a heat exchange \cite{Eli,Mohammady21} . Furthermore, interaction with the environment in stage (2) would induce dissipation within the system-apparatus complex, leading to entropy generation \cite{Popovic23,Francica21}. As such, the measurement process is subject to constraints stemming from the second \cite{Jacobs12,YiKim13,Guryanova19,Guryanova20,Strassberg22,Shettell22} and third \cite{Mohammady23,Taranto23} laws of Thermodynamics.

In this paper, our objective is to demonstrate that additional limitations on the measurement process arise not from Thermodynamics but rather from Relativity and Quantum Mechanics themselves. Specifically, we will focus on the fact that a successful quantum measurement entails recording the measurement outcome \cite{Hartle93,Hartle16,Hartle21,Barrett99,Barrett02}. This record is imprinted on a non interacting system whose states are persistent and distinguishable, usually a macroscopic system or a system with infinite degrees of freedom \cite{Hartle91}. The recording process consumes both energy and time, which must be considered when evaluating the efficiency and power of a quantum measurement engine.

For simplicity, our analysis will be based on a concrete example: a quantum Otto engine utilising a spin-1/2 particle as the working substance \cite{Talkner17}. This engine can be implemented both as a heat engine and a quantum measurement engine.

\begin{figure}[H]
\centering

\begin{subfigure}[b]{.32\linewidth}
\includegraphics[width=\linewidth]{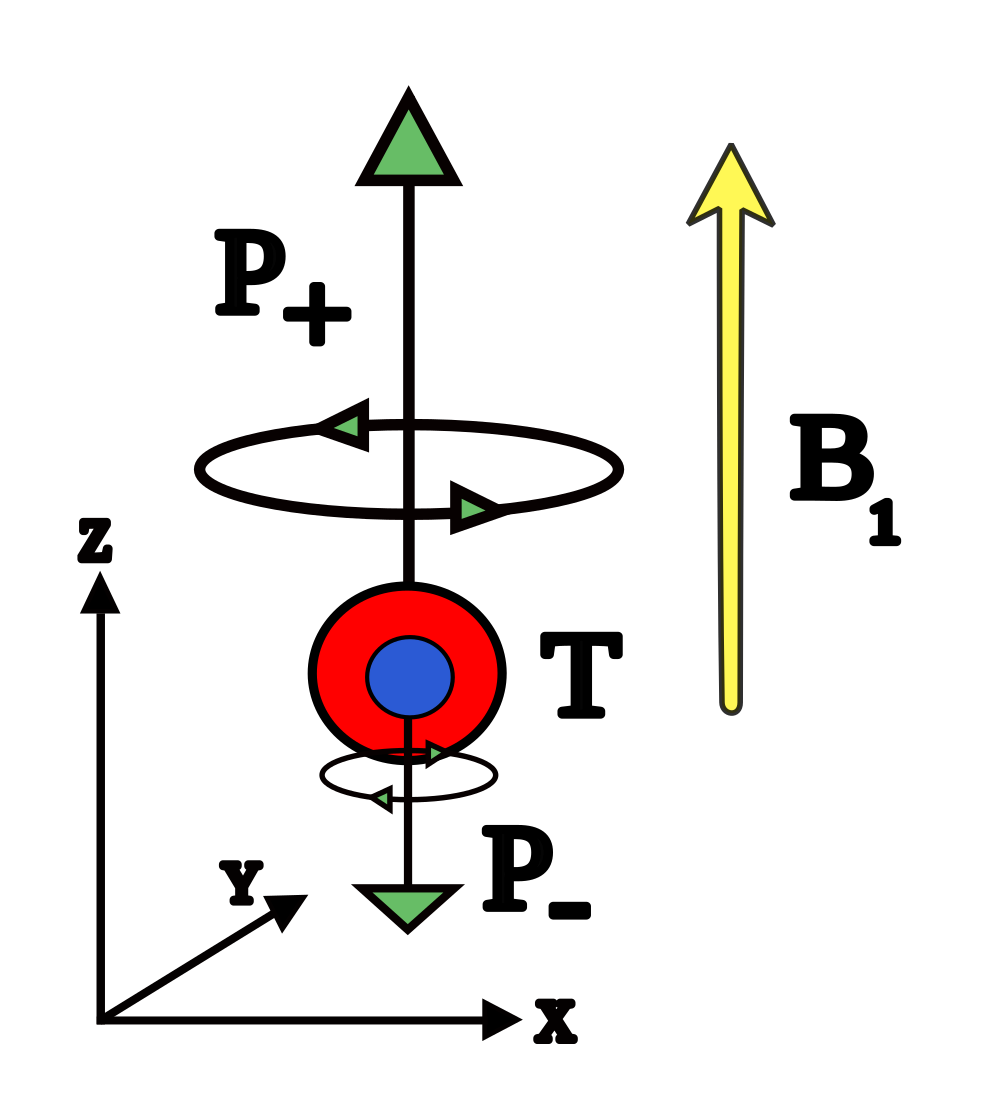}
\setcounter{subfigure}{0}%
\caption{}\label{Fig:Work_field}
\end{subfigure}
\begin{subfigure}[b]{.67\linewidth}
\includegraphics[width=\linewidth]{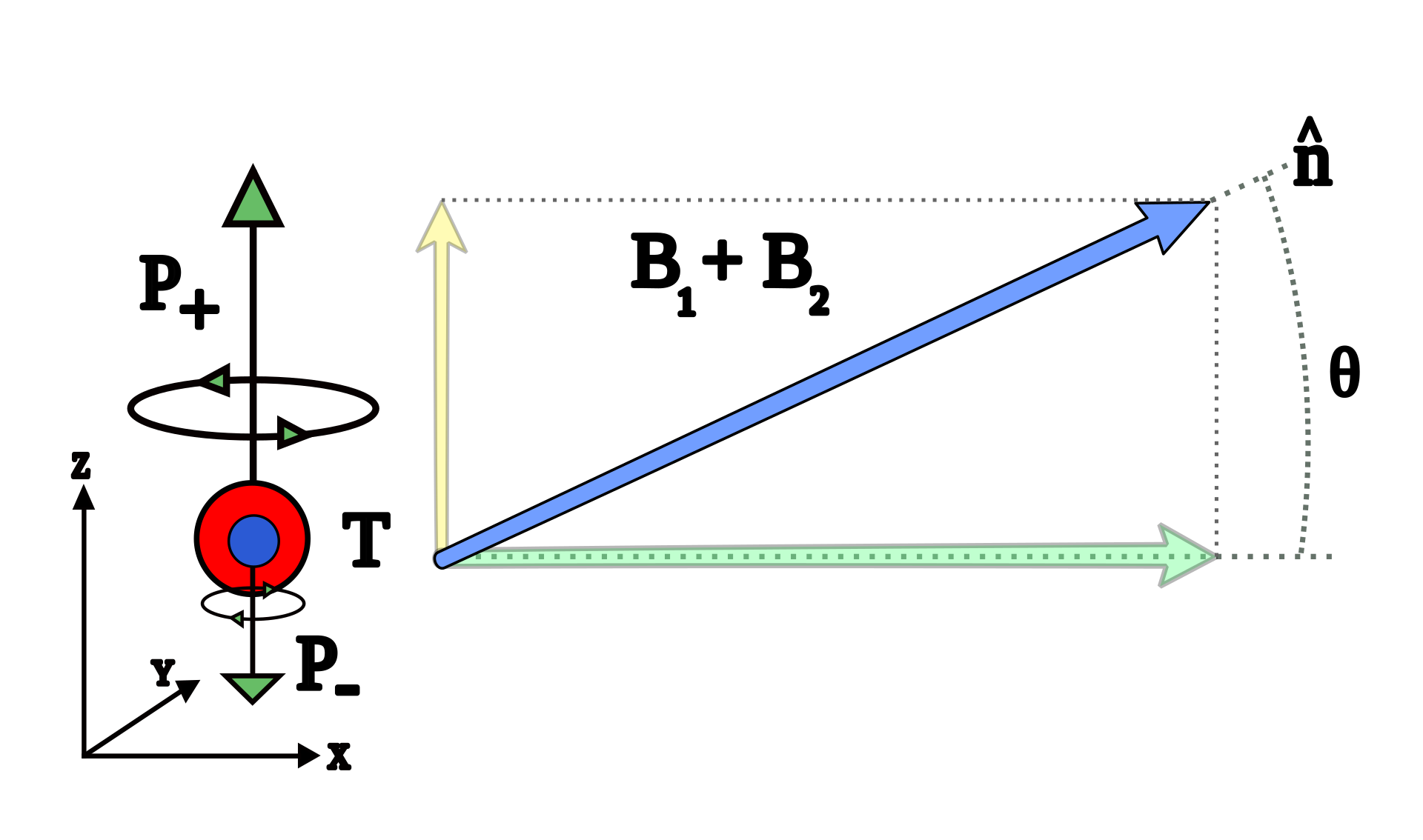}
\caption{}\label{Fig:Probe_field}
\end{subfigure}

\begin{subfigure}[b]{.45\linewidth}
\includegraphics[width=\linewidth]{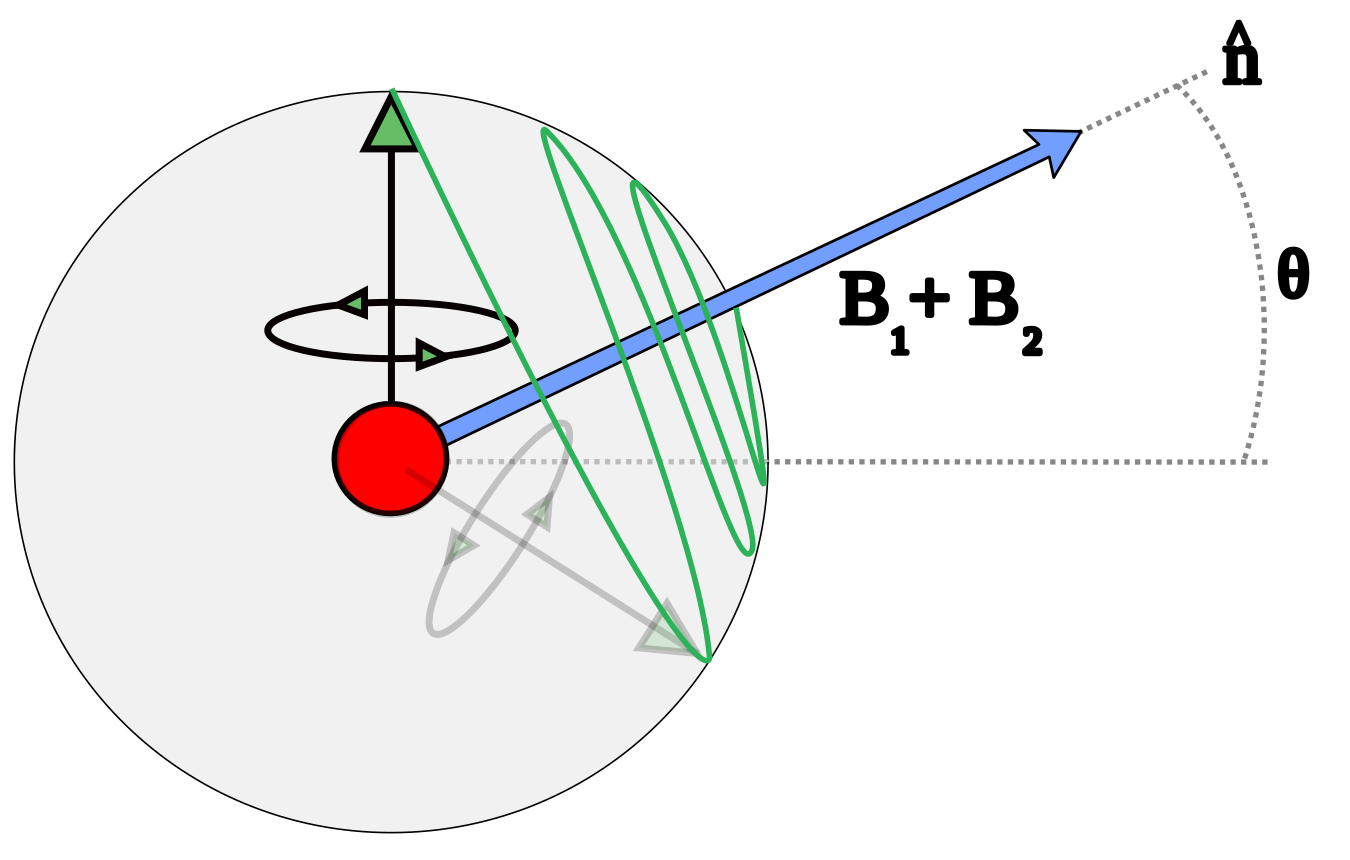}
\setcounter{subfigure}{2}%
\caption{}\label{Fig:Precesion}
\end{subfigure}

\caption{Schematics of the system. The blue and red dots with their respective thin arrows represent the spin 1/2 particle in its orthogonal states, their size being proportional to the probability of finding the system in that state. The thick arrows represent classical magnetic fields. a) State of the system after increasing the initial field from $B_0$ to $B_1$ in the $\hat{z}$ direction. b) Initial state after turning on the probe field. This represents the probe field implementation. c) Illustration of the spin precession and decay around the total field.}
\label{Fig:Scheme}
\end{figure}

In the heat engine implementation, the Otto cycle starts with the spin aligned with an external magnetic field of strength $B_0$ (see Fig. \ref{Fig:Work_field}). This field is then increased to $B_1$, causing the spin to perform work against the field (referred to as the work field). Subsequently, the spin is brought into contact with a high-temperature heat source, making both spin projections equally probable. At this temperature, reducing the work field back to $B_0$ incurs no energy cost. Finally, the spin is cooled back to the original state, statistically aligned with the work field. The efficiency of the Otto cycle is determined by $1 - \lambda$, where $\lambda = B_0 / B_1$ \cite{Talkner17}.

In the quantum measurement implementation, the hot bath is replaced by the measurement of the spin's projection on a direction orthogonal to the work field. Assuming the projection postulate, the spin collapses into a state oriented in the orthogonal plane, where both projections along the work field direction are equally probable. While the efficiency of the cycle would be the same as in the heat engine implementation if no additional heat exchanges related to the measurement were considered, it is generally accepted that at least the Landauer erasure heat should be included in the efficiency calculation.

Our goal is to analyze the measurement step more closely by proposing a specific measurement protocol rather than simply assuming a projective measurement. For simplicity we shall only consider this part of the machine cycle.

Our protocol involves applying a strong magnetic field of intensity $B_2$ (where $B_2 \ge B_1$) perpendicular to the original field, which we refer to as the probe field (see Fig. \ref{Fig:Probe_field}). 

Under the action of the magnetic fields, the spin precesses around the resultant of the work and probe fields \cite{GZ14,Eriksson2017} (see Fig. \ref{Fig:Precesion}). During precession, the time-dependence of the spin leads to the emission of electromagnetic waves. The actual amount of radiated energy depends on the original state of the spin, and so the final field state may be regarded as a record of the measurement outcome \cite{VK88}. The energy spent on building this record is extra cost of the working of the machine, and so it affects the machine efficiency. Similarly, the time spent on the spin relaxation must be taken into account in estimating the machine power. While other measurement protocols for two-dimensional Hilbert spaces, such as the Stern-Gerlach \cite{Stern-Gerlach} or the electron-shelving \cite{Wineland03} experiments, are easier to perform in a laboratory and are conceptually simpler, they usually rely on entangling intermediate degrees of freedom, such as the position of the particle or excited internal states, between the recording device and the actually measured subsystem, making the theoretical analysis less straightforward and most of the time only approximated or phenomenological. We believe the simplicity and direct interaction between the measured subsystem and the recording device (the radiated magnetic field) is a fundamental conceptual advantage for the proposed protocol.

The machine actually has two modes of operation. As a measurement device, it is best to allow the spin to get fully aligned with the work and probe fields, thus maximizing the radiated energy, and making the final states of the radiation field as different from each other as possible. However, this also makes for the lowest efficiency as an engine. So when regarded as an engine, the best strategy is to leave the probe field on only until the mean value of the spin projection along the work field vanishes, at which time the probe field is turned off. This is actually a short time in comparison to the spin relaxation time scale, so the amount of radiated energy is small. This makes for a poor measurement but an efficient engine, as we will show. The point is that while this may not be the usual way to measure a spin in practice, it nevertheless may be regarded as a paradigmatic measurement in so far as it contains the two essential elements of the measurement process: it  leaves the spin in a specific state based on the original value of its projection, and it leaves a record of this state, imprinted on a non interacting system with infinitely many degrees of freedom \cite{GS90}. Because of this, the system has been widely used as a paradigmatic measurement process and serves the purpose of analysing the dynamics in detail.

In the following, we solve for the joint evolution of the spin and the quantized radiation field by considering them as a single coupled quantum system \cite{VK88,LB39,Heisenberg1930,Jacobs06,Ramakrishna21,Auffeves22}, and thus find the final states of the radiation field, the energy taken from the spin, and the characteristic time of the process. The energy invested in the radiation field and the time required to generate it reduce both the engine's efficiency and power, and must be incorporated into their evaluation alongside the heat exchanges previously mentioned \cite{Jacobs12,Hofmann21,Talkner21,Landi22}. Although the actual effect is quantitatively small, we emphasise its fundamental character, as it follows essentially from the finiteness of $c$ and $\hbar$.

In summary, we contend that an effective quantum measurement requires the recording of its outcome. This is a physical process constrained by the principles of quantum mechanics and relativity, over and above other thermodynamic considerations, and has direct implications on the engine efficiency and power. 

This paper is structured as follows: In the following section, we introduce the Spin Quantum Otto cycle, fueled by a thermal source, which serves as our standard model. In section \ref{Sec:measurement}, we present our proposal and replace the thermal source with a quantum protocol that aims to mimic a measurement in specific conditions. Section \ref{Sec:eff} addresses the work, efficiency, and power of the engine. Finally, we offer some concluding remarks.

\section{A Spin Quantum Otto Cycle}
\label{Sec:Otto}
To initiate our discussion, we'll delve into the workings of a thermal engine, setting the stage for a comparative analysis with the forthcoming quantum measurement engine. Our focus will be on a straightforward realization of a spin quantum Otto cycle \cite{Talkner17}. It is worth noting that in \cite{Talkner17}, authors perform a projective measurement along the $\hat{x}$ axis, while on this work we replace this step by an interaction with an external field which will serve as measuring device as we will show. \\
Our system involves a spin-1/2 particle. At the cycle's inception, the particle resides in a thermal state at temperature $T$, interacting with a magnetic field $B_z(0)$ oriented along the $\hat{z}$-axis. This configuration implies a well-defined spin in the $z$ direction, where the spin assumes the value of 1/2 with a corresponding probability

\be
    p_+ = \frac{1}{1 + e^{-2 \beta_0 \mu B_z(0)}}
\te
and the value -1/2 with probability $p_- = 1- p_+$. $\beta_0$ is the initial inverse temperature multiplied by the Boltzmann constant and $\mu = {e \hbar}/{2m}$ with $e$ and $m$ the charge and mass of the particle respectively. The mean energy of the spin is
\be
    \langle E \rangle (0) = -\mu B_z(0) \left( p_+ - p_- \right).
\te

At the onset of the cycle's initial phase, we raise the field adiabatically to a new magnitude, denoted as $B_z(1)$, henceforth we shall call this field the ``work'' field. Throughout this process, entropy remains constant, while the mean energy experiences a reduction, yielding:
\be
    \langle E \rangle (1) = -\mu B_z(1) \left( p_+ - p_- \right).
\te
Hence, the machine yields work
\be
    W_{01} = - \left[ \langle E \rangle (1) - \langle E \rangle (0) \right] = \mu \left[B_z(1) - B_z(0) \right] \left( p_+ - p_- \right)
\te

During the second leg, we let the spin evolve to a state where its orientation along the x-axis becomes well-defined, assuming either value with equal probability. This alignment can be achieved by coupling the system to a heat bath at infinite temperature. Regardless of the approach, the heat exchange incurred is irreversible, as the system and bath are at different temperatures. The resulting mean energy is now
\be
    \langle E \rangle (2) = 0
\te
and the exchanged heat is
\be
    Q_{12} = \langle E \rangle (2) - \langle E \rangle (1) = \mu B_z(1) \left( p_+ - p_- \right)
\te
During the third stage, we return the field adiabatically to its initial value, $B_z(0)$. As $p_+$ equals $p_-$ throughout this process, there is no net exchange of work. Subsequently, we enable the system to undergo thermalization once more, releasing a heat
\be
    Q_{30} = - \langle E \rangle (0)
\te

It is natural to write down the efficiency
\be
    \eta_O = \frac{W_{01}}{Q_{12}} = 1 - \frac{\langle E \rangle (0)}{\langle E \rangle (1)} := 1 - \lambda
\te
where 
\be
\label{Eq:lambda}
    \lambda = \frac{B_z(0)}{B_z(1)}
\te

\section{Implementation as a quantum measurement engine}
\label{Sec:measurement}

The goal of implementing the quantum Otto cycle \cite{Talkner17} as a quantum measurement engine is to avoid energy exchanges under the form of heat, thus sidestepping limitations imposed by the second law of Thermodynamics.

In the thermal engine implementation of the Otto cycle, the main heat exchange happens in the second leg, where the two projections of the spin along the work field are brought to equiprobability. A natural replacement for this operation is to force the spin to be projected upon the orthogonal plane. According to the usual projection postulate \cite{VN31}, this may be realized by a spin measurement along any direction on that plane.

Our strategy will be somewhat different as we shall consider instead a destructive measurement of the \emph{original} spin projection along the work field, but with the same final outcome of leaving the spin projected along the orthogonal plane. 

We shall force the spin to evolve by applying a probe field of strenght $B_2$ in the $\hat x$ direction. The direction of the total field is $\hat{n}$ (work plus probe fields, see Fig. \ref{Fig:Probe_field}). The probe field will be kept on until the mean value of the spin projection along the work field vanishes, at which point the probe field is turned off.

Our goal is to show that in the process a small amount of energy is radiated as electromagnetic waves. The actual final state of the field depends on whether the spin was initially up or down, and so it may be regarded as a record of the original spin projection. 

To do this we shall not appeal to the projection postulate but rather solve for the joint evolution of the spin and the radiation fields, a strategy inspired by Heisenberg's analysis of the bubble chamber in \cite{Heisenberg1930}.  For simplicity we shall restrict this analysis to the second leg of the cycle only, assuming that the other legs proceed as in the usual implementation. 

\subsection{The model}
We assume the work and probe fields are classical, constant fields along the duration of the second leg. \\

The radiation field is expanded in modes, leading to the quantized magnetic field
\be
\label{Eq:quant_EM_field}
\vec{B}_q = i \sqrt{\frac{\hbar}{\epsilon_0 c}} \int \frac{DK}{\sqrt{2k}} \left[ \left( \vec{k} \times \vec{e}_{\mu} \right) a_{K} e^{i \vec{k}\cdot\vec{r}} - \left( \vec{k} \times \vec{e}^{\dagger}_{\mu} \right) a_{K}^{\dagger} e^{-i \vec{k}\cdot\vec{r}} \right]
\te
$\epsilon_0$ is the vacuum permittivity. The modes are indexed by $K=\left(\bm{k},\alpha\right)$ where $\bm{k}$ is the momentum and $\alpha$ denotes the polarization vector $\epsilon_{\alpha}$. 
Here

\be
\int\;DK=\int\frac{d^3k}{\left(2\pi\right)^3}\sum_{\alpha}
\te
$a_{K}$ and $a_K^{\dagger}$ are the creation and destruction operators for the corresponding mode. They have units of $k^{-\frac{3}{2}}$. 

The spin is carried by a particle with magnetic moment $\mu = \frac{q \hbar}{2 m}$, the magnetic moment of the particle (it is assumed that $g = 2$ and $q$ is the charge of the particle), is so that $\vec{\mu} \cdot \vec{B}$ has energy units. Since the particle is effectively far from the field sources, the interaction takes a dipole-dipole form and can be written as $-\vec{\mu}\cdot\vec{B}$ \cite{GZ14, Eriksson2017, LB39}. This particle is further assumed to be captured into a harmonic trap, so that its position is a Gaussian variable whose uncertainty $\sigma$ is the smallest length scale in the problem.\\ 

The spin-radiation field complex evolves under the Hamiltonian

\be
H = H_S\otimes\bm{1}_{ph}+\bm{1}_{S}\otimes H_{ph} + H_I 
\te
where $H_S$ is the Hamiltonian for the spin under the classical magnetic fields

\be
H_S = -\hbar\Omega \sigma_{\hat{n}} 
\te 
The spin operator along the $\hat n$ direction of the total classical magnetic field will be represented by the Pauli matrix $\sigma_3$, and $\hbar\Omega=\mu\sqrt{B_1^2+B_2^2}$, $B_1$ and $B_2$ being the magnitudes of the work ($B_1$) and probe ($B_2$) fields.\\ 

$H_{ph}$ is the free Hamiltonian for the radiation field

\be
H_{ph} = \int\;DK\;\hbar\omega_Ka_K^{\dagger}a_K
\te 
and $H_I$ is the interaction Hamiltonian

\be
H_I = \sigma_+ \otimes B^{\dagger} + \sigma_- \otimes B
\te
where $\sigma_{\pm}$ are the raising and lowering spin operators along the direction $\hat n$, while the field operator $B$ is

\be 
B=-i\int\;DK\;\bm{K}_{K-}a_K 
\te
Here, the interaction Hamiltonian is an approximation of the term $-\vec{\mu} \cdot \vec{B}_q$ coming from (\ref{Eq:quant_EM_field}), where we dropped the longitudinal term (associated with the $\hat{n}$ direction) assuming that the quantized contribution to the total field in this direction is negligible compared to the classical one. If we were working in the interaction picture, this could be regarded as a rotating wave approximation, where the highly improbable spin-flips against the classical field are neglected. \\
To define the amplitude $\bm{K}_{K-}$ we first introduce the vector 

\be 
\bm{K}_K=\mu \sqrt{\frac{\hbar}{\epsilon_0 c}} F\left[k\right]\left(\frac{\bm{k}\times\epsilon_{\alpha}}{\sqrt{2k}}\right)
\te 
where 

\be
\label{Eq:regularization} 
F[k] = e^{-\frac{k^2 \sigma^2}{2}}
\te 
is a structure function which takes into account the uncertainty in the spin localization. We then project the vector $\bm{K}_K$ on the plane perpendicular to $\hat n$, where we choose two cartesian coordinates $1$ and $2$, and define

\be
\bm{K}_{K+} =\bm{K}^*_{K-} = \bm{K}_{K1} - i\bm{K}_{K2} 
\te

\subsection{Dynamics}

The system of spin plus radiation field is described by a density matrix $\rho$ obeying the Liouville- von Neumann equation

\be
i\hbar\frac d{dt}\rho = [H,\rho]
\label{LvN}
\te
The spin and field separately are described by the Landau traces $\rho_S={\rm{Tr}}_{ph}\rho$, $\rho_{ph}={\rm{Tr}}_{S}\rho$. We write \cite{Lax1964,Sargent1978}

\be 
\rho=\rho_S\otimes\rho_{ph} + \rho_{c}
\label{Lax}
\te
$\rho_c$ describes the entanglement between both subsystems; both its partial traces vanish ${\rm{Tr}}_{ph}\rho_c={\rm{Tr}}_{S}\rho_c=0$. Introducing the decomposition (\ref{Lax}) in (\ref{LvN}) and taking the corresponding partial traces we get 

\be
\label{SEq}
i\hbar \dot{\rho}_S = [H_S, \rho_S] + {\rm{Tr}}_{ph} [H_I, \rho_S\otimes\rho_{ph}] + {\rm{Tr}}_{ph} [H_I, \rho_{c} ]
\te 

\be
\label{PHEq}
i\hbar \dot{\rho}_{ph} = [H_{ph}, \rho_{ph}] + {\rm{Tr}}_{S}[H_I, \rho_S\otimes\rho_{ph}] + {\rm{Tr}}_S [H_I, \rho_{c}] 
\te
We have used that ${\rm{Tr}}_S [H_{ph}, \rho_{c}]= [H_{ph}, {\rm{Tr}}_S\;\rho_{c}]=0$, and also ${\rm{Tr}}_{ph} [H_{S}, \rho_{c}]=0$. For the correlation part of the system we get
\be
\label{CEq}
i \hbar\dot{\rho}_{c} =  [H, \rho_{c}] + [H_I, \rho_s\otimes\rho_{ph}] - {\rm{Tr}}_{ph}\left( [H_I, \rho] \right) \otimes\rho_{ph} - \rho_S \otimes {\rm{Tr}}_S\left( [H_I, \rho] \right)
\te
Without loss of generality we may parameterize 

\be
\rho_S=\frac{1}{2} \left[\bm{1} + \vec r\cdot\vec \sigma\right] = \frac{1}{2} \left[\bm{1} + r_3\sigma_{\hat{n}} + r_+ \sigma_+ + r_- \sigma_-\right]
\label{param}
\te 
where $r_-=r_+^*$ and $r_3^2+r_+r_-\le 1$, with equality for a pure state.

Our strategy will be to find an approximate expression for $\rho_{ph}$ and $\rho_c$ for an arbitrary evolution of the $r_{\pm}$ and $r_3$ parameters, to first order in the interaction Hamiltonian, which can then be introduced into the equation for $\rho_S$ (\ref{SEq}) to obtain an equation valid to second order in the interaction, to be solved self-consistently.

\subsection{Radiation field dynamics}

We assume an uncorrelated initial state so that $\rho_c$ is itself of first order in the interaction. Then to find the field density matrix to first order we may neglect $\rho_c$. Introducing the parameterized $\rho_S$ from eq. (\ref{param}) into eq. (\ref{PHEq}) we obtain

\be 
i \hbar\dot{\rho}_{ph} = [H^{eff}_{ph}, \rho_{ph}]
\te 
where

\be
H_{ph}^{eff}=\int\;DK\;h_{K}^{eff}
\te
and 

\be
h_{K}^{eff}=\hbar\omega_Ka_K^{\dagger}a_K+\frac  i2  \left( r_-\bm{K}_{K+}a_K^{\dagger}-r_+\bm{K}_{K-}a_K\right)
\te
If the initial state of the field is the vaccum, then it evolves  into a normalized coherent state

\be 
a_K\left|z_K\right\rangle=z_K\left|z_K\right\rangle
\label{Radiation}
\te
where

\be
z_K=\frac{1}{2\hbar} \bm{K}_{K+}\int_0^tdt'\;e^{-i\omega_K\left(t-t'\right)}r_-\left(t'\right)
\te

\subsection{Correlation dynamics}

Using $\rho_S$ as in eq. (\ref{param}) and the first order reduced density matrix for the radiation field 

\be 
\rho_{ph}\approx \prod_{K}\;\left|z_K\right\rangle \left\langle z_K\right| 
\label{FOph}
\te 
into eq. (\ref{CEq}) we get for the correlation density matrix

\be
\label{coherent}
i \hbar \dot{\rho}_{c} = [H_0, \rho_{c}] +F\left(t\right)
\te
with $H_0=H_S + H_{ph}$, and 

\be 
F=\sigma_{\hat{n}}\otimes f_3 +\sigma_+\otimes f_+ +\sigma_-\otimes f_- 
\te 
where the $f_j$ are photonic operators,  $f_3=-f_3^{\dagger}$ and $f_-=-f_+^{\dagger}$; explicitly

\bea
f_3&=&-\frac14r_3\left( r_-\left[B^{\dagger},\rho_{ph}\right]+r_+\left[B,\rho_{ph}\right]\right)+\frac14 \left(r_- \left\{B^{\dagger},\rho_{ph}\right\}-r_+ \left\{B,\rho_{ph}\right\}\right)\nn
&+&\frac{\hbar}2 \left(r_- \Omega_{eff}-r_+ \Omega_{eff}^*\right)\rho_{ph}
\tea

\be
f_+=-\frac14r_+\left( r_-\left[B^{\dagger},\rho_{ph}\right]+r_+\left[B,\rho_{ph}\right]\right)-\frac{1}2r_3 \left(\left\{B^{\dagger},\rho_{ph}\right\}+2\hbar\Omega_{eff}\rho_{ph}\right)+\frac12\left[B^{\dagger},\rho_{ph}\right]  
\te
where 

\be
\hbar\Omega_{eff}=-i\int\;DK\;\bm{K}_{K+}z^*_K
\te
The solution is

\be
\label{FOc}
\rho_c = -i \int_{0}^{t}\;dt'\;e^{-i H_0 \left( t- t'\right)} F(t') e^{i H_0 \left( t- t' \right)}
\te

\subsection{Spin dynamics} 

To obtain the spin dynamics we use the first order $\rho_{ph}$ from eq. (\ref{FOph}) and $\rho_c$ from eq. (\ref{FOc}) into the equation (\ref{SEq}) for the spin reduced density matrix. We also define

\bea
r_{\pm}&=&e^{\pm i2\Omega t}\bar r_{\pm} \nn
r_{\hat{n}}&=&1-r_3
\tea

Then
\be
\dot{r}_{\hat{n}} = - \frac12 \int_o^t dt'\; H(t-t') \left[ \bar r_+(t') \left( \bar r_-(t) - \bar r_-(t') \right) + r_{\hat{n}}(t') \right] + C.C.
\label{one}
\te
and 
\be
\dot{\bar r}_+ = -\int_0^t\;dt'\;H\left(t-t'\right)\bar{r}_+\left(t'\right) \left[1+r_{\hat{n}}(t')e^{-2i\Omega\left(t-t'\right)}-r_{\hat{n}}(t)\right]
\label{two}
\te
where

\be
H\left(t-t'\right)=\frac{1}{\hbar^2}\int\;DK\;\bm{K}_{K+}\bm{K}_{K-}e^{-i\left( 2\Omega-\omega_K\right) \left( t-t'\right) }
\te
The dynamics generated by eqs. (\ref{one}) and (\ref{two}) is similar to the one encountered in a quantum mechanical decay problem, namely at very early times $r_z$ and $\bar r_{\pm}$ are quadratic in time, then turn into an exponential decay and finally a power law approach to the final equilibrium state \cite{Peres80,Razavy13}. Careful consideration shows that the quadratic period is so short that it can be neglected without loss of accuracy (see Appendix), and by the time of the final approach to equilibrium spin and radiation are already effectively decoupled. Moreover we shall assume that initially the probability of the spin being up is dominant. Then $\left| r_{\pm}\right|$ is never large and we may linearize equations (\ref{one}) and (\ref{two}) around the stable equilibrium at $r_z=r_{\pm}=0$  

The decay time may be found as an approximate pole in the Laplace transform of $r_z$ and $\bar{r}_+$

\be 
\left[ s-\frac{i}{\hbar^2}\int\;DK\;\frac{\bm{K}_{K+}\bm{K}_{K-}}{2\Omega-\omega_K-is}\right] \bar r_+\left( s\right) =\bar r_+\left( t\approx 0\right) 
\te 
where the right hand side is taken at some point at the beginning of the exponential stage. There is indeed an approximate pole at 

\be 
s\approx -\xi + i\phi
\te 
with 

\bea
\label{Eq:xi_phi}
\xi &=& \frac{\pi}{\hbar^2}\int\;DK\;\bm{K}_{K+}\bm{K}_{K-}\delta\left( 2\Omega-\omega_K\right) \nn
\phi &=& \hbar^{-2}P.V.\left( \int \frac{DK\; \bm K_{K+} \bm K_{K-}}{\omega_K - 2\Omega - \phi} \right) \approx 0
\tea
As $\xi$ is positive, equation (\ref{Eq:xi_phi}) ensures stability in the linear regime.  While on the exponential stage we may approximate

\bea
\label{Eq:r_approx}
\bar r_{\pm} &\approx& \bar r_{\pm}^{0}e^{ -\xi t} \nn
r_{\hat{n}} &\approx& r_z^{0}e^{ -\xi t}
\tea
The value of $\xi$ eq. (\ref{Eq:xi_phi}) is the essential input we need to analyze the machine efficiency and power.

\section{Analysis}
\label{Sec:eff}
\subsection{The engine cycle}

Initially, the spin points in the $+\hat{z}$ direction with probability $p_+$ and the are no photons. The initial density matrix is then
\be
    \rho_0 = \frac{1}{2} \begin{pmatrix}
1 & \Delta \\
\Delta & 1
\end{pmatrix} \otimes |0\rangle\langle0|
\te
where $\Delta = p_+ - p_- = \tanh\left( \beta \mu B \right)$ with $\beta = \left( k_B T \right)^{-1}$. The probe field $B_2$ is applied in the $x$ direction. Then, the total magnetic field points in the $\hat{n} := \cos \theta \hat{x} + \sin \theta \hat{z}$ direction. In order to compute the radiation due to the spin precession, we will work with this rotated axis and assume that the $\sigma_3$ Pauli matrix is associated with the $\hat{n}$ direction. 
As the working system we consider only the spin and therefore only its pure energy (and not the interaction one) will be taken into account for thermodynamic purposes. That is to say
\be
E_s = \langle H_S\rangle_{\rho_S}
\te
We define $\hbar \Omega_i \equiv \mu B_i$. The engine cycle is as follows:

\begin{itemize}
    \item In the first step we extract work by making the magnetic field grow from $B_0$ to $B_1$ in the $\hat{z}$ direction. The extracted work is
    \be
    \label{Eq:work1}
    W_1 = \Delta \hbar\left( \Omega_1 - \Omega_0 \right)
    \te
    \item The motivation for the next step is to be able to decrease this bigger field by making no work in the system. For that, we turn on a probe magnetic field, $B_2$,  in the $\hat{x}$ direction. This field will make the spin precede and radiate. Turning this field on costs no energy at all because of the initial condition. It is important to note that in this step the system will radiate.
    \item Now the spin has a certain alignment with the $\hat{n}$ direction and turning $B_2$ off will cost a work
    \be
    E_{off} = \hbar\Omega_2 \cos \left( \theta \right) r_3 - \hbar\Omega_1 \cos \left( \theta \right) Re\left( r_+ \right)
    \te
    Where $r_3$ and $r_{\pm}$ correspond to the evolution in the $\hat{n}$ axis and $\tan\left(\theta\right) = \frac{B_1}{B_2}$. \\
    The better aligned the spin and the total field are, the more expensive the energetic price to turn the probe field off.
    \item Now that the spin is (imperfectly) aligned in the $\hat{n}$ direction, decreasing the work field in the $\hat{z}$ direction from $B_1$ to $B_0$ requires a work exchange, depending on the alignment direction. If the projection along the $\hat{z}$ axis is positive, work must be fed into the system. Otherwise, a negative projection would allow for an useful work extraction:
    \be
    \label{Eq:work2}
    W_2 = -\hbar\left[ \Omega_1 - \Omega_0 \right] \left[ \cos\left( \theta \right) Re\left( r_+ \right) + \sin\left( \theta \right) r_3 \right]
    \te
    This term must be taken into account as work (either positive or negative) as it is a coherent change in the $\hat{z}$ magnetic field, which is the exact way we extracted work in the first step.
    \item In this last step, we let the spin thermalize again. This releases a heat
    \be
    Q = \hbar\Omega_0 \left[ \Delta - \cos\left( \theta \right) Re\left( r_+ \right) - \sin\left( \theta \right) r_3 \right]
    \te
\end{itemize}

\subsection{Efficiency}
The efficiency is 
\be
\eta = \frac{W}{E_{off}} = \frac{\left[ \Omega_1 - \Omega_0 \right] \left[ \Delta - \Delta_0 \right]}{\cos\theta\left[ \Omega_2 r_3 - \Omega_1 Re\left( r_+ \right) \right]}
\te
where we have defined $\Delta_0 := \cos\left( \theta \right) Re\left( r_+ \right) + \sin\left( \theta \right) r_3$ and $W := W_1 + W_2$. \\
In order to compare with previous results, the strategy we will adopt is to turn the $B_2$ field off when the mean value of the spin is in the $x-y$ plane, so $Re\left(r_+\right) = -r_3 \tan \theta = -r_3 \frac{\Omega_1}{\Omega_2}$ and $\Delta_0(t_c) = 0$.
Then, for $t=t_c$ we have
\be
\eta_c = \frac{\left[\Omega_1 - \Omega_0\right] \Delta}{\cos\theta\ r_3\left(t_c\right)\left[ \frac{\Omega_1^2 + \Omega_2^2}{\Omega_2} \right]} = \frac{\left[ \Omega_1 - \Omega_0 \right] \Delta}{\cos\theta\ r_3(t_c) \frac{\Omega^2}{\Omega_2}
}
\te
Recall that $\Omega^2 = \Omega_1^2 + \Omega_2^2$. 
To make the machine work we need to have $\Omega_2 > \Omega_1 > \Omega_0$. To study the efficiency we parameterise
\bea
\label{Eq:gamma}
\lambda \Omega_1 &=& \Omega_0 \nn
\gamma \Omega_2 &=& \Omega_1
\tea
We arrive to the formula:
\be
\label{Eq:eta_c}
\eta_c = \left[ 1 - \lambda \right] \left[ \frac{\Delta \sin\theta}{r_3(t_c)} \right] = \left[ 1 - \lambda \right] \left[ \frac{r_3(0)}{r_3(t_c)} \right] = \left[ 1 - \lambda \right] \left[ \frac{1-r_{\hat{n}}\left(0\right)}{1-r_{\hat{n}}\left(0\right)e^{-\xi t_c}} \right]
\te
In this expression it is clear that, if $\xi = 0$, we recover the original result from \cite{Talkner17}.

The fact that $t_c$ is finite also places limits on the power which may be extracted from the engine. We shall not analyse this issue in detail because the end result is consistent with expectations from a simple ``quantum speed limit'' analysis \cite{Mandelstam45,Lloyd03,Deffner17}.

\subsection{Numerical results}
\begin{figure}[H]
    \centering
    \includegraphics[scale=0.8]{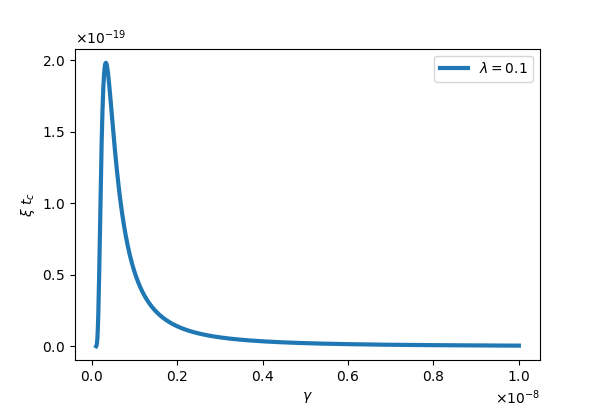}
    \caption{$\xi\ t_c$ as a function of the parameter $\gamma$ (which measures the probe field strength, see Eq. (\ref{Eq:gamma})). The efficiency is a decreasing function of this quantity. Here we show the regime where the radiation correction is maximum, where the probe field is several orders of magnitude bigger than the work field. For $\gamma \to 1$ this quantity is always decreasing.}
    \label{Fig:graph_1}
\end{figure}

\begin{figure}[H]
    \centering
    \includegraphics[scale=0.7]{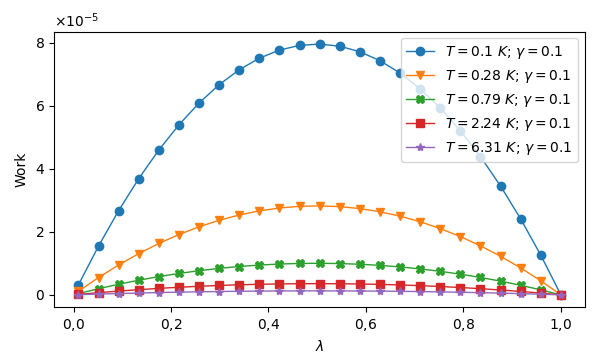}
    \caption{Dimensionless work $W_{dim}$ (normalised by the characteristic energy of the system $\mu B_1$), eq. (\ref{Eq:dimensionless_work}), as a function of the $\lambda$ parameter (which measures the work field strength, see Eq. (\ref{Eq:lambda})).}
    \label{Fig:graph_2}
\end{figure}

\begin{figure}[H]
    \centering
    \includegraphics[scale=0.7]{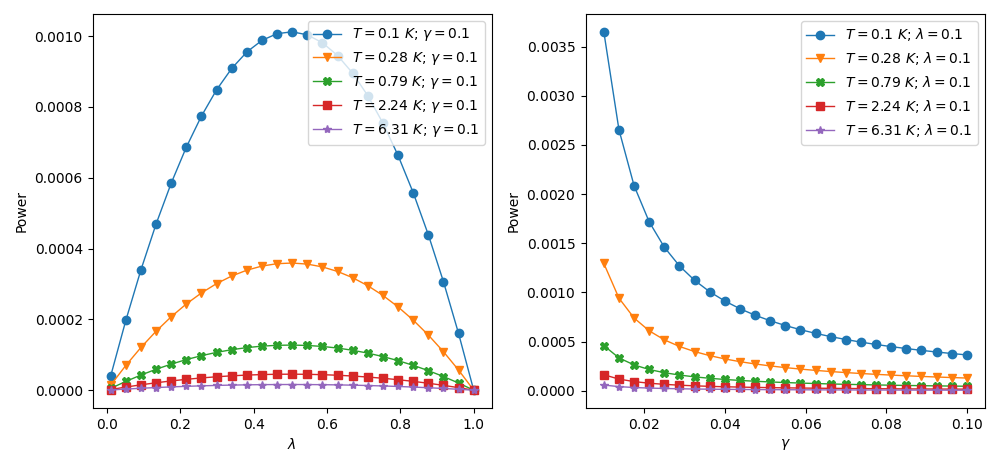}
    \caption{Dimensionless power $P_{dim}$ (normalised by the characteristic power of the system $\frac{\left(\mu B_1\right)^2}{\hbar}$), eq. (\ref{Eq:dimensionless_power}), as a function of the $\lambda$ and $\gamma$ parameters, which parameterise the work and probe fields strength respectively.}
    \label{Fig:graph_3}
\end{figure}

We assume that the particle is in a harmonic trap of frequency $\Omega_{trap}$ so $F\left[ k \right]$ in (\ref{Eq:regularization}) is
\be
F(k) = e^{- \frac{1}{2} \sigma^2 k^2}
\te
with $\sigma = \sqrt{\frac{\hbar}{m \Omega_{trap}}}$. \\
For the critic time, when we turn off the probe field, we have the transcendental equation
\be
Re\left(r_+(t)\right) = -\tan \theta \; r_3(t)
\te
which, consistently with the accuracy of the other computations, we approximate by
\be
    t_c \approx \frac{\arccos \left( - \gamma^2 \right)}{2 \Omega}
\te
where we put the initial conditions $r_{\pm}(0) = \Delta \cos \theta$ and $r_3(0) = \Delta \sin \theta$. \\
We recall that
\bea
r_3(t) &\approx& 1 - \left( 1 - \Delta \sin \theta \right) e^{-\xi t} \nn
r_{\pm} &\approx& \Delta \cos\left( \theta \right) e^{\left( -\xi \pm i 2\Omega \right) t} \nn
\Delta &=& \tanh\left( \beta \hbar\Omega_0\right) \nn
\beta &=& \left( k T \right)^{-1}
\tea
and report the following parameters for the numerical results
\bea
m &=& 2000\ m_e \nn
q &=& q_e \nn
B_1 &=& 0.1\  \rm Tesla \nn
\Omega_{trap} &=& 100 \ \Omega_1
\tea
Where $m_e$ and $q_e$ are the electron mass and charge respectively. \\
The numerically quantities studied are $\xi\ t_c$, of which efficiency is a decreasing function given by eq. (\ref{Eq:eta_c}) (see Figure \ref{Fig:graph_1}), the dimensionless work
\be
\label{Eq:dimensionless_work}
W_{dim} = \frac{W_1 + W_2}{\mu B_1},
\te
eqs. (\ref{Eq:work1}) and (\ref{Eq:work2}) normalised by the characteristic energy of the system $\mu B_1$ (see Figure \ref{Fig:graph_2}), and the dimensionless power
\be
\label{Eq:dimensionless_power}
P_{dim} = \frac{W}{t_c} \frac{\hbar}{\left( \mu B_1 \right)^2}
\te
(see Figure \ref{Fig:graph_3}).
The radiation correction to the efficiency is peaked when the probe field is tuned with the characteristic frequency of the particle width. The power is a decreasing function of $\gamma$ and we observe a trade-off in $\lambda$. When $\lambda \to 1$ there is no change in the initial field and no work is extracted but when $\lambda \to 0$, $p_+ = p_-$ and the thermal state has equal probabilities in the $\hat{z}$ direction, so increasing the work field produces no work.

\section{Final remarks}
From a thermodynamic perspective, a machine in contact with a single thermal reservoir is impossible, and one in contact with two has, at most, the Carnot efficiency. When the energy is not supplied by a thermal source, the Carnot bound ceases to hold, but still other limitations arise. Assuming the projection postulate it has been claimed that the simple act of measuring a device can fuel it in a very efficient way \cite{Elouard17, Elouard18, Laflamme2020, Auffeves21, Auffeves21c, Goold23} but a more realistic measurement model is very much needed to shed light on this issue. Quantum measurement involves a complex interaction between the object system, the measuring apparatus, a recording device and the environment at large. Further, in any realistic case, the measurement apparatus must have a large number of degrees of freedom \cite{VK88}.\\
Our model gives us an scenario where this complexity may be fully explored. It yields to a first principle description, the fuelling dynamics can be fully studied, a measurement limit of this system can be taken and it shows very clearly that, when a ``to be measured'' quantum system, i.e., the spin 1/2 particle in this case, comes into contact with a ``measurement device'' (a macroscopic object, i.e., the quantized electromagnetic field with uncountable many degrees of freedom), the energy spreads out and both efficiency and power  decrease. 

Even though in the case at hand this phenomenon is quantitatively rather small, and the losses in efficiency and power can be minimised varying the parameter of the model, the point is that limitations on efficiency, power, and measurement accuracy are linked in such a way that a certain compromise is unavoidable: an improvement on any of these implies a loss on the others. For this reason the particulars of the measurement process must be taken into account in order to make the model of the machine complete.  \\

Similar analyses must be done in the field of Quantum Measurement Based Quantum Computers \cite{Wei21, Deffner21}. The impossibility for a quantum measurement engine to do work in a perfectly efficient way shows that the finiteness of $\hbar$ and $c$ lead to limitations in the feasibility of certain physical processes that resemble those coming from the laws of thermodynamics, even in regimes where these do not apply, at least directly.\\

\section{Acknowledgments}

Work supported in part by  Universidad de Buenos Aires through Grant No. UBACYT
20020170100129BA, CONICET Grant No. PIP2017/19:11220170100817CO and ANPCyT Grant
No. PICT 2018: 03684.

\appendix
\section{Analysis for early times}
For short times we use the ansatz $r_z(t) = r_z(0) e^{-f(t)}$, $\bar{r}_+(t) = \bar{r}_+(0) e^{-g(t)}$ and write the Taylor series in time around $t=0$ for $f$, $g$ and $H$:
\bea
f(t) &=& a_0 + a_1 t + \frac12 a_2 t^2 + \frac16 a_3 t^3 + ... \nn
g(t) &=& b_0 + b_1 t + \frac12 b_2 t^2 + \frac16 b_3 t^3 + ... \nn
H(t) &=& H_0 + H_1 t + \frac12 H_2 t^2 + \frac16 H_3 t^3 + ...
\tea

By doing so we compute
\bea
a_0 &=& 0 \ \ \ , \ \ \ b_0 = 0 \nn
a_1 &=& 0 \ \ \ , \ \ \ b_1 = 0 \nn
a_2 &=& H_0 \ \ \ , \ \ \ b_2 = H_0\ \nn
a_3 &=& \frac12 \left[H_1 + H_1^*\right] = 0 \ \ \ , \ \ \ b_3 = - \left[ H_1 + H_0 r_z\left(0\right) 2i\Omega \right]
\tea
As $H_0$ is real and positive defined, this tells us that in the regime of applicability of our machine both variables tend to decrease and, after a short period, the system tends to its linear regime.

In order to match the early time regime to the exponential decay given by eqs. (\ref{Eq:r_approx}), we search for a time $t_*$ and an initial condition $r_z(t_*)$ such that both solutions and their first derivatives coincide:
\bea
r_z(t_*) e^{-\xi t_*} &=& r_z(0) e^{-\frac{H_0}{2} t_*^2} \nn
-\xi r_z(t_*) e^{-\xi t_*} &=& - H_0 t_* r_z(0) e^{-\frac{H_0}{2} t_*^2}
\tea
From where we find that
\bea
t_* &=& \frac{\xi}{H_0} \nn
r_z(t_*) &=& r_z(0) e^{\frac{\xi^2}{2H_0}}
\tea
This tells us that, in our regime where $\xi \ll \Omega$ and $H_0 \sim \sigma^{-4} \gg \left(\frac{m\Omega}{\hbar}\right)^2$, the exponential dominates from very early times and it is a very good approximation to simply set the initial conditions for this function.

\end{document}